\documentclass[manuscript]{aastex}
\usepackage{spr-astr-addons}
\usepackage{graphicx}
\usepackage[dvips]{epsfig}
\makeatletter
\renewcommand*{\@biblabel}[1]{\hfill(#1) }
\makeatother
\begin{document}

\title{ Gravitational waves from r-mode instability of massive young sub- and super-Chandrasekhar white dwarfs }

\author{Somnath Mukhopadhyay$^1$ and Sarmistha Banik$^1$}

\affil{$^1$BITS Pilani, Hyderabad Campus,\\ Telangana - 500078 , India.\\
somnathm@hyderabad.bits-pilani.ac.in\\
sarmistha.banik@hyderabad.bits-pilani.ac.in}

\maketitle


\begin{abstract}
    In the present work we investigate the r-mode instability windows, spindown and spindown rates of sub- and super-Chandrasekhar magnetized white dwarfs in presence of Landau quantization of the electron gas and magnetic braking. The gravitational wave strain amplitudes due to r-mode instability is also calculated. The dominant damping mechanism is taken to be the shear viscosity arising due to scattering of the degenerate electrons from the ion liquid. We find that the critical frequencies of Landau quantized magnetized white dwarfs are the lowest, those of non-Landau quantized ones are higher and those of non-magnetized ones are the highest at the same temperature. This implies that magnetic braking and Landau quantization both enhance r-mode instability. We have also seen that there is rapid spindown of magnetized white dwarfs due to additional magnetic braking term but there is no considerable effect of Landau quantization on the spindown and spindown rates for magnetic field strengths relevant for white dwarf interiors. We find that the r-mode gravitational wave strain amplitude for a rapidly rotating super-Chandrasekhar white dwarf at 1 kpc is $\sim 10^{-27}$, making isolated massive rapidly rotating hot magnetized white dwarfs prime candidates for search of gravitational waves in the future.
\end{abstract}



\noindent
\section{Introduction}
\label{Section 1} 
    
    Radial and non-radial pulsations of general relativistic compact stellar models are important for asteroseismology due to their intimate connection to gravitational pulsations. The corresponding general relativistic analysis of non-radial perturbation over hydrostatic equilibrium using Regge-Wheeler choice of gauge for non-rotating or slowly rotating spherically symmetric star can be found in \citep{Th67}. These perturbations are also called quasinormal modes. Depending upon the nature of the perturbations these modes are given various names. One of the most important modes in a compact star is the Rossby mode or R-mode, which is driven unstable by the emission of gravitational waves via the Chandrasekhar-Friedman-Schutz (CFS) mechanism due to the flow pattern being retrograde in the frame of the star and prograde for an observer at infinity \citep{Ch1970,Fr1978}. The damping of r-modes is caused by dissipative processes inside the star such as bulk and shear viscosities, where the properties of dense matter or the role of the Equation of State (EoS) comes into play. Hence detecting gravitational wave strain amplitudes from quasinormal modes of white dwarfs and neutron stars in the future by detectors such as advanced LIGO, advanced Virgo, Kagra, Einstein telescopes etc. will not only shed light on the nature of superdense matter but also on the nature of gravity as well \citep{Gla2018}. 
    
    Recent studies \citep{Das12,Das13,Ko14,Be14,Ch17,Mu17} have shown the existence of magnetized white dwarfs with masses much larger than the traditional Chandrasekhar limit \citep{Ch35},also known as Super-Chandrasekhar white dwarfs to account for the exceptionally high luminosities of the Type Ia supernovae, e.g. SN2003fg, SN2006gz, SN2007if, SN2009dc  \citep{Ho06,Sc10,Hi07,Ya09,Si11}. In these studies, the electron gas in the EoS is taken to be free, relativistic and Landau quantized in a strong magnetic field. Coulomb corrections to the EoS, magnetization, instabilities due to general relativity, pycnonuclear and electron capture reactions are considered in \citep{Ch17} and general relativistic magnetostatic equilibrium models were considered in \citep{Ko14,Be14}. There were other recent studies on the possiblity of Soft Gamma-ray Repeaters (SGRs) and Anomalous X-ray Pulsars (AXPs) as strongly magnetized white dwarfs \citep{Ma12}. About $\sim$250 magnetic white dwarfs with observed surface magnetic fields have been observed \citep{Fe15}. If white dwarfs with no or uncertain field determination are also included the number can reach $\sim$600 \citep{Ke13,Ke15}. The Sloan Digital Sky Survey (SDSS), Hamburg Quasar Survey (HQS) and the Cape Survey show that the distribution of surface magnetic field strengths of magnetized white dwarfs lies in the range 10$^3$-10$^9$ gauss. However, the internal magnetic field strength, which is unobserved, can be many orders of magnitude greater than the surface field. In fact, the central magnetic field is the key factor in determining the stability of super-Chandrasekhar magnetized white dwarfs. However, at a very high central density, electron capture and pycnonuclear reactions in the core can lead to softening of the EoS which can further lead to local instabilities as said previously. These factors can constrain the upper limit to the central magnetic field strength of the white dwarf, which typically ranges from 10$^{14}$-10$^{16}$ gauss depending on the core composition.
    
    Although the growth time of r-modes in white dwarfs is very long ($\sim 10^8$ years) \citep{An1999}, it may be excited in massive, rapidly spinning, accreting white dwarfs. In the present study we explore the r-mode instability windows, spindown and spindown rates of hot, rapidly rotating sub- and super-Chandrasekhar white dwarfs in presence of magnetic braking and Landau quantization of the electron gas. We also calculate the corresponding gravitational wave strain amplitudes for r-mode instability. 
     
\noindent
\section{Formalism}
\label{Section 2}
\subsection{Gravitational, Magnetic and Viscous timescales} 
\label{subsection 2.1}
 
    We determine the r-mode instability windows in presence of shear viscosity due to electron-ion scattering and strong magnetic fields in rapidly rotating hot, ultramagnetized sub- and super-Chandrasekhar white dwarfs. Hence, it is necessary to consider the effects of radiation on the evolution of mode energy. This is expressed as the integral of the fluid perturbation \citep{Lindblom1998,Lindblom1999},
    
\begin{equation}
\widetilde{E}=\frac{1}{2}\int{\left[ \rho \delta \vec{v}.\delta \vec{v}^{*}+\left(\frac{\delta p}{\rho}-\delta \Phi \right)\delta \rho^{*}\right]}d^{3}r,
\label{eq1}
\end{equation}
where $\rho$ is the mean density profile of the star, $\delta \vec{v}$, $\delta p$, $\delta \Phi$ and $\delta \rho $ are perturbations in the velocity, pressure, gravitational potential and density due to oscillation of the mode respectively. The dissipative timescale of an r-mode is \citep{Lindblom1998},

\begin{equation}
\frac{1}{\tau_{i}}=-\frac{1}{2\widetilde{E}}\left(\frac{d\widetilde{E}}{dt}\right)_{i},
\label{eq2}
\end{equation}
where, the index '$i$' refers to the various dissipative mechanisms, i.e., gravitational wave emissions, viscosity and magnetic braking.

To the lowest order of $\delta \vec{v}$ and $\delta \rho$ the expression for energy of the mode in Eq.(1) can be reduced to a one-dimensional integral \citep{Lindblom1998,Vidana2012} in the small angular velocity limit 

\begin{equation}
\widetilde{E}=\frac{1}{2}\alpha_r^{2} R^{-2l+2} \Omega^{2} \int^{R}_{0} \rho(r) r^{2l+2} dr, 
\label{eq3}
\end{equation}
where, R is the radius of the white dwarf, $\alpha_r$ is the dimensionless amplitude of the mode, $\Omega$ is the angular velocity of the white dwarf and $\rho(r)$ is the radial dependence of the mass density. Since the expression of $(\frac{d\widetilde{E}}{dt})$ due to gravitational radiation \citep{Thorne1980,Owen1998}, shear viscosity \citep {2Lindblom1999} and magnetic braking \citep{Ho2000} are known, Eq.(2) can be used to evaluate the imaginary part of $\frac{1}{\tau}$. It can be written as

\begin{equation}
\frac{1}{\tau(\Omega,T)}=\frac{1}{\tau_{GR}(\Omega,T)}+\frac{1}{2\tau_{M}(\Omega,T)}+\frac{1}{\tau_{SV}(\Omega,T)},
\label{eq4}
\end{equation}
where $1/\tau_{GR}$, $1/\tau_{M}$ and $1/\tau_{SV}$ are the contributions from gravitational radiation, magnetic braking and shear viscosity, respectively, and are given by

\begin{eqnarray}
\frac{1}{\tau_{GR}}=-\frac{32 \pi G \Omega^{2l+2}}{c^{2l+3}} \frac{(l-1)^{2l}}{[(2l+1)!!]^2}\left(\frac{l+2}{l+1}\right)^{(2l+2)}\nonumber \\
 \times\int^{R}_{0}\rho(r)r^{2l+2} dr, 
\label{eq5}
\end{eqnarray}
\begin{equation}
\frac{1}{\tau_{M}}=\frac{B_s^2R^6\Omega^2}{6c^3I},
\label{eq6}
\end{equation}
\begin{eqnarray}
\frac{1}{\tau_{SV}}=(l-1)(2l+1)\int_{0}^{R}\eta r^{2l}dr\left(\int_{0}^{R}\rho r^{2l+2}dr\right)^{-1}, 
\label{eq7}
\end{eqnarray}
where, G and c  in Eq.(5) are the gravitational constant and velocity of light; $B_s$ and $I$ in Eq.(6) are the surface magnetic field and moment of inertia of the white dwarf and $\eta$ and $l$ in Eq.(7) are the shear viscosity coefficient and mode index respectively.

    The dominant contribution to the shear viscosity in hot white dwarf interiors is from the scattering of electrons with the ion liquid. The fit to the shear viscosity coefficient is given by \citep{2Lindblom1999, Na84}

\begin{equation}
\eta=\frac{10^6\rho_6^{2/3}}{1+1.62\rho_6^{2/3}I_2}, 
\label{eq8}
\end{equation}
where,
\begin{eqnarray}
I_2=0.667log(1.32+0.103T_6^{1/2}\rho_6^{-1/6})\nonumber \\
+0.611-\frac{0.475+1.12\rho_6^{2/3}}{1+1.62\rho_6^{2/3}}, 
\label{eq9}
\end{eqnarray}
where all the quantities are given in CGS units and $T_6$ and $\rho_6$ are measured in $10^6$ K and $10^6$ $g cm^{-3}$. In order to have transparent visualisation of the role of angular velocity and temperature on various timescales, it is useful to factor them out by defining fiducial timescales. Thus, we define fiducial gravitational timescale $\widetilde{\tau}_{GR}$ such that \citep{Lindblom1998,Lindblom2000},

 
\begin{equation}
\tau_{GR}=\widetilde{\tau}_{GR} \left(\frac{\Omega_0}{\Omega}\right)^{2l+2},
\label{eq11}
\end{equation}
and the fiducial magnetic braking timescale is given by,

\begin{equation}
\tau_{M}=\widetilde{\tau}_{M} \left(\frac{\Omega_0}{\Omega}\right)^2,
\label{eq12}
\end{equation}
where $\Omega_0=\sqrt{ \pi G \bar{\rho}}$ and $\bar{\rho}= 3M/4 \pi R^3$ is the mean density of white dwarf having mass $M$ and radius $R$.
Thus Eq.(4) becomes

\begin{equation}
\frac{1}{\tau(\Omega,T)}=\frac{1}{\widetilde{\tau}_{GR}}\left(\frac{\Omega}{\Omega_0}\right)^{2l+2}+\frac{1}{2\widetilde{\tau}_{M}}\left(\frac{\Omega}{\Omega_0}\right)^2+\frac{1}{\tau_{SV}}.
\label{eq13}
\end{equation}

    For a given temperature and mode $l$ the equation for critical angular velocity $\Omega_c$ is obtained from the condition $\frac{1}{\tau(\Omega_c,T)}=0$. At a given T and mode $l$, the equation for the critical velocity is a polynomial of order $l+1$ in $\Omega_c^{2}$ and thus each mode has its own characteristic $\Omega_c$. Since the smallest of these, i.e. $l=2$, is the dominant contributor, study is being done for this mode only. The critical angular velocity $\Omega_c$ in terms of $\Omega_0$ for this mode is obtained from the equation
    
\begin{equation}
\frac{1}{|\widetilde{\tau}_{GR}|}\left(\frac{\Omega_c}{\Omega_0}\right)^6+\frac{1}{2\widetilde{\tau}_{M}}\left(\frac{\Omega_c}{\Omega_0}\right)^2-\frac{1}{\tau_{SV}}=0
\label{eq14}
\end{equation}

Once the Equation of State (EoS) is ascertained, then all physical quantities necessary for the calculation of r-mode instability can be performed.
\noindent
\subsection{Spindown and Spindown Rates}
\label{subsection 2.2}

When the angular velocity of a white dwarf goes beyond the critical value $\Omega_c$, the instability of r-mode sets in and the star emits gravitational radiation which takes away the energy and angular momentum, resulting the star to spin down to the region of stability. The evolution of the angular velocity when the r-mode amplitude $\alpha_r$ reaches saturation ($d\alpha_r/dt=0$), as the angular momentum is radiated to infinity by gravitational radiation, is given by \citep{Owen1998,Mu18}

\begin{equation}
\frac{d\Omega}{dt}=\frac{2\Omega}{\tau_{GR}}\frac{\alpha_r^2Q}{1-\alpha_r^2Q}.
\label{eq15}
\end{equation}
\noindent
If, in addition to gravitational radiation, electromagnetic radiation due to magnetic braking is considered, then Eq.(14) changes to \citep{Ho2000}

\begin{equation}
\frac{d\Omega}{dt}=\frac{2\Omega}{\tau_{GR}}\frac{\alpha_r^2Q}{1-\alpha_r^2Q}-\frac{\Omega}{\tau_{M}}\frac{1}{1-\alpha_r^2Q},
\label{eq16}
\end{equation}  
where $Q=\frac{3\tilde{J}}{2\tilde{I}}$ and

\begin{equation}
 \tilde{J}=\frac{1}{MR^4}\int_0^R{\rho(r)r^6dr}
 \label{eq17}
\end{equation}
and
\begin{equation}
 \tilde{I}=\frac{8\pi}{3MR^2}\int_0^R{\rho(r)r^4dr}
 \label{eq18}
\end{equation}
\noindent
We can see from Eq.(15) that the first term on the right hand side scales as $\Omega^7$ (since $|\tau_{GR}|^{-1} \propto \Omega^6$) and the second term scales as $B_s^2\Omega^3$ (since $\tau_{M}^{-1} \propto B_s^2\Omega^2$).

\noindent
\subsection{ Equation of State for non-magnetized White Dwarfs }
\label{subsection 2.3} 

    We take a completely degenerate relativistic free electron gas at absolute zero. Due to Pauli's exclusion principle, the electrons encounter a degeneracy pressure which keeps them moving and hence the total energy of the Fermi gas is greater than the single-electron ground state energies. Since this degeneracy pressure is non-zero even when the temperature is zero it is sufficient to stabilize a white dwarf star (a Fermi gas of electrons) against gravitational collapse.
    
    The degeneracy pressure of the electron gas is given by    
    
\begin{equation}
 P_e= \frac{1}{3}\int p v n_p d^3p = \frac{1}{3}\int \frac{p^2c^2}{\sqrt{(p^2c^2+m_e^2c^4)}} n_p d^3p
\label{eq19}
\end{equation}
where $m_e$ is the electron rest mass, $v$ is the velocity of the electrons with momentum $\vec p$ and $n_p d^3p$ is the  electron number density with momenta between $\vec p$ and $\vec p + d\vec p$. The factor $\frac{1}{3}$ accounts for the spatial isotropy of pressure. For electrons having spin $\frac{1}{2}$, degeneracy = 2, $n_p d^3p = \frac{8\pi p^2 dp}{h^3}$ and hence number density $n_e$ is given by 

\begin{equation}
 n_e= \int_0^{p_F} n_p d^3p = \frac{8\pi p_F^3}{3h^3} = \frac{x_F^3}{3\pi^2\lambda_e^3}
\label{eq20}
\end{equation}
where $p_F$ is the Fermi momentum of the electron gas, $x_F=\frac{p_F}{m_e c}$ is the dimensionless Fermi momentum and $\lambda_e=\frac{\hbar}{m_e c}$ is the electron Compton wavelength. The energy density of the electron gas $\varepsilon_e$  is given by

\begin{equation}
 \varepsilon_e= \int_0^{p_F} E n_p d^3p = \int_0^{p_F} \sqrt{(p^2c^2+m_e^2c^4)} \frac{8\pi p^2 dp}{h^3}
\label{eq21}
\end{equation}
which can be integrated along with Eq.(19) to give

\begin{equation}
 \varepsilon_e= \frac{m_e c^2}{\lambda_e^3} \chi(x_F); ~~~~  P_e= \frac{m_e c^2}{\lambda_e^3} \phi(x_F),
\label{eq22}
\end{equation}
where

\begin{equation}
 \chi(x)= \frac{1}{8\pi^2}[x\sqrt{1+x^2}(1+2x^2)-\ln(x+\sqrt{1+x^2})]
\label{eq23}
\end{equation}
and

\begin{equation}
 \phi(x)= \frac{1}{8\pi^2}[x\sqrt{1+x^2}(\frac{2x^2}{3}-1)+\ln(x+\sqrt{1+x^2})].
\label{eq24}
\end{equation}
\noindent
To calculate the EoS for non-magnetized white dwarfs, pressure $P$ is given by 

\begin{equation}
P = P_e= \frac{m_e c^2}{\lambda_e^3} \phi(x_F),
\label{eq25}
\end{equation}
and for the total energy density $\varepsilon$ both electrons (with their kinetic and rest mass energies) and rest mass energies of atomic nuclei contribute, so that 
    
\begin{eqnarray}
\varepsilon= \varepsilon_e + n_e(m_p+f m_n)c^2\nonumber\\
 =\frac{m_e c^2}{\lambda_e^3} \chi(x_F) + n_e(m_p+f m_n)c^2. 
\label{eq26}
\end{eqnarray}
Here $m_n$ and $m_p$ are the neutron and proton masses, respectively and $f$ is the number of neutrons per electron. Atomic nuclei contributes to the mass of white dwarf and the pressure is provided by the degenerate electron gas. Usually, white dwarf stars consist of helium, carbon, oxygen, etc., for which the mass number is approximately twice the atomic number and hence $f=1$. To be more accurate, one should also subtract $n_e(1+f)$ times the binding energy per nucleon from the rest mass energy of atomic nucleus. This binding energy correction term depends on the composition and its effects are insignificant, e.g. in case of a helium white dwarf star it is about 0.7$\%$ of the rest mass of nucleus, and hence it is ignored in the calculations. 

\noindent
\subsection{Landau quantization and Equation of State for magnetized White Dwarfs }
\label{subsection 2.4}

    Like the previous case, here also we consider a completely degenerate free relativistic electron gas at zero temperature but immersed in a strong background magnetic field. Electrons, being charged, are now Landau quantized. Landau quantization modifies the EoS, which, in turn, changes the thermodynamic quantities of the electron gas like pressure and energy density. The magnetic field energy density and pressure are also taken into consideration. The total matter and magnetic field pressure and energy density determines the stability and mass-radius relationships of  ultramagnetized white dwarfs. It is to be noted that at the values of magnetic field strengths relevant for white dwarf interiors, protons are weakly Landau quantized. This is because the mass of proton being $\sim 1836$ times larger than the mass of electron the proton cyclotron energy is $\sim 1836$ times lesser than that of the electron for the same value of the magnetic field, and hence it is ignored.

    In order to find the thermodynamic quantities of a magnetized electron gas at zero temperature, we need to calculate the energy spectrum and the density of states. The quantum dynamics of a charged particle in a constant magnetic field is described in many previous articles (e.g. Sokolov and Ternov (1968) \citep{ST68}, Landau and Lifshitz (1977) \citep{LL77}, Canuto and Ventura (1977) \citep{CV77}  M\'esz\'aros (1992) \citep{Me92}). We assume a uniform magnetic field $B$ directed along the positive z-axis and consider the dynamics of a charged particle of charge $q$ and mass $m_e$. In classical electrodynamics, the particle moves in a circular trajectory with radius and angular frequency (cyclotron frequency) given by
    
\begin{equation}
 r_c =\frac{m_ecv_{\perp}}{qB};    ~~~~    \omega_c=\frac{qB}{m_ec}
\label{eq27}
\end{equation}    
where $v_{\perp}$ is the velocity of the particle transverse to the magnetic field direction. The Hamiltonian of the system is given by  

\begin{equation}
H =\frac{1}{2m_e} \Big(\vec p - \frac{q\vec A}{c}\Big)^2
\label{eq28}
\end{equation} 
where $\vec B = \nabla \times \vec A$ and $\vec A$ is the vector potential. The magnetic vector potential is given by

\begin{equation}
 \vec A = \bordermatrix{ &\cr 
                        &0 \cr
                        &Bx \cr
                        &0\cr}                      
\label{eq29}
\end{equation}
and therefore

\begin{equation}
H =\frac{1}{2m_e} [p_x^2 + \Big(p_y - \frac{qBx}{c}\Big)^2 + p_z^2]
\label{eq30}
\end{equation}
Since the operator $y$ is absent, the operator $\hat p_y$ commutes with this Hamiltonian . Hence $\hat p_y$ can be replaced by its eigenvalue $\hbar k_y$. Using the expression for cyclotron frequency $\omega_c=\frac{qB}{m_ec}$ we get 

\begin{equation}
H =\frac{p_x^2}{2m_e} + \frac{1}{2}m_e\omega_c^2\Big(x - \frac{\hbar k_y}{m_e\omega_c}\Big)^2 + \frac{p_z^2}{2m_e}.
\label{eq31}
\end{equation}
The first two terms in Eq. (30) depict the Hamiltonian of a quantum harmonic oscillator with the potential minimum shifted in coordinate space by $x_0=\frac{\hbar k_y}{m_e\omega_c}$. Since coordinate translation of the oscillator potential leaves the energies unaffected, the energy eigenvalues can be written as

\begin{equation}
E_{n ,p_z} =(n+\frac{1}{2})\hbar \omega_c + \frac{p_z^2}{2m_e}, ~~~~n=0, 1, 2 . . . . 
\label{eq32}
\end{equation}
The set of quantum states with a particular value of the quantum number $n$ is called a Landau Level. Each Landau level is degenerate as the energy is independent of $k_y$. If periodic boundary condition is assumed $k_y$ can take values $k_y=\frac{2\pi N}{l_y}$ where $N$ is another integer and $l_x,l_y,l_z$ are the dimensions of the system considered. The values of $N$ are constrained by the fact that the centre of the oscillator force $x_0$ must physically lie within the system, $0\le x_0 \le l_x$ which implies $0\le N \le \frac{l_x l_y m_e\omega_c }{2\pi\hbar}=\frac{qB l_x l_y}{hc}$. Applying this to electrons having charge $q=-|e|$ and spin $s$, the maximum number of particles in each Landau level per unit area is $\frac{|e|B(2s+1)}{hc}$. If we solve Schr\"odinger's equation with z-component of spin angular momentum $s_z$ in a constant and uniform magnetic field in z-direction, then Eq.(31) modifies to 

\begin{equation}
E_{\nu ,p_z}=\nu\hbar \omega_c + \frac{p_z^2}{2m_e}, ~~~~ \nu=n+\frac{1}{2}+s_z.
\label{eq33}
\end{equation}
The spin degeneracy $g_\nu=1$ for the lowest Landau level ($\nu=0$) and $g_\nu=2$ (for $s_z=\pm\frac{1}{2}$) for higher Landau levels ($\nu\neq0$).

    For extremely strong magnetic fields such that $\hbar \omega_c \geq m_ec^2$ electrons become relativistic. In that case, solving the Dirac equation in a constant magnetic field \citep{Lai91} gives the energy eigenvalues
    
\begin{equation}
E_{\nu ,p_z}=\left[p_z^2c^2+m_e^2c^4\left(1+2\nu B_d\right)\right]^\frac{1}{2}
\label{eq34}
\end{equation}    
where the dimensionless magnetic field defined as $B_d=B/B_c$ is introduced with $B_c$ given by $\hbar\omega_c=\hbar\frac{|e|B_c}{m_ec}=m_ec^2 \Rightarrow B_c=\frac{m_e^2c^3}{|e|\hbar}=4.414\times 10^{13}$ gauss. The density of states at zero temperature can be written as

\begin{equation}
\sum\limits_{\nu }\frac{2|e|B}{hc}g_{\nu}\int \frac{dp_z}{h},
\label{eq35}
\end{equation}
and hence the electronic number density is given by

\begin{equation}
n_e=\sum\limits_{\nu =0}^{\nu_m} \frac{2|e|B}{h^2c} g_{\nu} \int_0^{p_F(\nu )} dp_z = \sum\limits_{\nu =0}^{\nu_m}\frac{2|e|B}{h^2c}g_{\nu}p_F(\nu)
\label{eq36}
\end{equation}
where $p_F(\nu )$ is the Fermi momentum in the $\nu $th Landau level and $\nu_m$ is the highest Landau level index. The electron Fermi energy $E_F$ in the $\nu $th Landau level is given by

\begin{equation}
E_F^2=p_F^2(\nu)c^2+m_e^2c^4\left(1+2\nu B_d\right)
\label{eq37}
\end{equation}
and $\nu_m$ can be found from the condition $[p_F(\nu)]^2 \geq 0$ or

\begin{equation}
\nu_m =\frac{\epsilon_{F max}^2-1}{2B_d},
\label{eq38}
\end{equation}     
where $\epsilon_{F max}=\frac{E_{F max}}{m_ec^2}$ is the dimensionless maximum Fermi energy of electrons for fixed $B_d$ and $\nu_m$. Very weak magnetic fields ($B_d<<1$) give very large number of Landau levels which then forms a continuum. The maximum value of Landau level index, $\nu_m$, is taken to be the nearest lowest integer. In terms of a dimensionless Fermi momentum $x_F(\nu)=\frac{p_F(\nu )}{m_ec}$, Eqns.(35) and (36) may be written as

\begin{equation}
n_e=\frac{2B_d}{(2\pi )^2\lambda_e^3}\sum\limits_{\nu =0}^{\nu_m} g_{\nu}x_F(\nu )
\label{eq39}
\end{equation}
and

\begin{equation}
\epsilon_F=\left[x^2_F(\nu )+1+2\nu B_d\right]^\frac{1}{2}
\label{eq40}
\end{equation}
or

\begin{equation}
x_F(\nu )=\left[\epsilon_F^2-(1+2\nu B_d)\right]^\frac{1}{2}.
\label{eq41}
\end{equation}   
The electron energy density is given by

\begin{eqnarray}
\varepsilon_e=\frac{2B_d}{(2\pi )^2\lambda_e^3}\sum\limits_{\nu =0}^{\nu_m} g_{\nu }\int_0^{x_F(\nu )}E_{\nu ,p_z}d\left(\frac{p_z}{m_ec}\right) \nonumber\\
=\frac{2B_d}{(2\pi )^2\lambda_e^3}m_ec^2\sum\limits_{\nu =0}^{\nu_m} g_{\nu }(1+2\nu B_d)\nonumber\\
\times \psi \left(\frac{x_F(\nu )}{(1+2\nu B_d)^{1/2}}\right),
\label{eq42}
\end{eqnarray}
where

\begin{eqnarray}
\psi (z)=\int_0^z(1+y^2)^{1/2}dy   \nonumber\\
=\frac{1}{2}[z\sqrt{1+z^2}+\ln(z+\sqrt{1+z^2})]
\label{eq43}
\end{eqnarray}
The degeneracy pressure of the magnetized electron gas is given by

\begin{eqnarray}
P_e=n_e^2\frac{d}{dn_e}\left(\frac{\varepsilon_e}{n_e}\right)= n_eE_F -\varepsilon_e \nonumber\\
=\frac{2B_d}{(2\pi )^2\lambda_e^3}m_ec^2\sum\limits_{\nu =0}^{\nu_m} g_{\nu }(1+2\nu B_d)\nonumber\\
\times\eta \left(\frac{x_F(\nu )}{(1+2\nu B_d)^{1/2}}\right)
\label{eq44}
\end{eqnarray}
where

\begin{eqnarray}
\eta (z)=z\sqrt{1+z^2}-\psi (z)    \nonumber\\
=\frac{1}{2}[z\sqrt{1+z^2}-\ln(z+\sqrt{1+z^2})].
\label{eq45}
\end{eqnarray}
\noindent
For the EoS of magnetized white dwarfs, the magnetic field energy density $\varepsilon_B=\frac{B^2}{8\pi}$ and pressure $P_B=\frac{1}{3}\varepsilon_B$ should be combined to the matter energy density and pressure as 
    
\begin{eqnarray}
P=&&P_e+P_B  \nonumber\\
=&&\frac{2B_d}{(2\pi )^2\lambda_e^3}m_ec^2\sum\limits_{\nu =0}^{\nu_m} g_{\nu }(1+2\nu B_d)\eta \left(\frac{x_F(\nu )}{(1+2\nu B_d)^{1/2}}\right)  \nonumber\\
&&+\frac{B^2}{24\pi},
\label{eq46}
\end{eqnarray}
and
  
\begin{eqnarray}
\varepsilon=&& \varepsilon_e + n_e(m_p+f m_n)c^2+\varepsilon_B   \nonumber\\
=&&\frac{2B_d}{(2\pi )^2\lambda_e^3}m_ec^2\sum\limits_{\nu =0}^{\nu_m} g_{\nu }(1+2\nu B_d)\psi \left(\frac{x_F(\nu )}{(1+2\nu B_d)^{1/2}}\right)  \nonumber\\
&&+ n_e(m_p+f m_n)c^2  +\frac{B^2}{8\pi}.
\label{eq47}
\end{eqnarray}

\noindent
\subsection{Constraining r-mode amplitudes from thermally equilibrated isolated white dwarfs}
\label{subsection 2.5}

    We consider white dwarfs to be in thermal equilibrium. In the thermal steady state, the gravitational radiation pumps energy into the r-mode at a rate given by \citep{Ma13,Mu19}
    
\begin{equation}
W_d=\frac{1}{3}\Omega J_c=-\frac{2\widetilde{E}}{\tau_{GR}}
\label{eq48}
\end{equation}
where
\begin{equation}
J_c=-\frac{3}{2}\tilde{J}MR^2\Omega \alpha_r^2,
\label{eq49}
\end{equation}
is the canonical angular momentum of the r-mode.
\noindent    
We further assume that all of the energy emitted from the star is due to the r-mode dissipation inside the star which implies \citep{Ma13} 
\begin{equation}
W_d=L_\gamma=4\pi R^2\sigma T_{eff}^4
\label{eq50}
\end{equation}    
where $\sigma$ is Stefan's constant, $T_{eff}$ is the effective surface temperature and $L_\gamma$ is the thermal photon luminosity at the surface of the star. From Eqs.(47),(48) and (49), we get
\begin{equation}
\alpha_r=\frac{5\times3^4}{2^8\tilde{J}MR^3\Omega^4}\left(\frac{L_\gamma}{2\pi G}\right)^{1/2}.
\label{eq51}
\end{equation}

\noindent     
\section{Results and Discussions} 
\label{Section 3}
\subsection{Critical Frequencies, Spindown and Spindown rates of sub- and super-Chandrasekhar white dwarfs}
\label{subsection 3.1}

    We calculate the masses and radii of non-magnetized and magnetized white dwarfs using the Tolman-\\Oppenheimer-Volkoff (TOV) equations. If one considers a very high central density for $f=1$ white dwarfs, one can asymptotically reach the Chandrasekhar mass limit. Beyond a density of $\sim 4.3\times10^{11}$ $g cm^{-3}$, the neutron drip point \citep{Ch15}, neutron rich nuclei appear and with further increase of density the free neutron states start populating, and the phase consists of neutron rich nuclei in a lattice embedded in a sea of electron and neutron gas.
    
    For magnetized white dwarfs we take the EoS with and without Landau quantization of electron gas. Instead of a constant magnetic field we take a density-dependent magnetic field profile to be consistent. We take the variation of magnetic field \citep{Ba97} in the interior of a magnetized white dwarf to be  
    
\begin{equation}
 B_d = B_{sd} + B_0[1-\exp\{-\beta(n_e/n_0)^\gamma\}]
 \label{eq52}
\end{equation} 
where $B_d$ is the dimensionless magnetic field at electron number density $n_e$, $B_{sd}$ is the dimensionless surface magnetic field, $n_0$ is the central electron number density ($n_e$(r=0)) and $B_0$, $\beta$ and $\gamma$ are constants. $B_0$ can be computed from Eq. (51) by fixing the surface and central magnetic field values. Constants $\beta=0.8$ and $\gamma=0.9$ are chosen to provide stable magnetized white dwarf equilibrium models. Once central and surface magnetic fields are fixed, the variations of its profile in the stellar interior do not affect the results significantly. Moreover, the maximum value of the magnetic field at the centre is kept at 10$B_c$ which is $4.414 \times 10^{14}$ gauss, lower than the maximum limit prescribed by N. Chamel et al. \citep{Ch13} and surface magnetic field strength $\sim10^{9}$ gauss determined from observational data \citep{Fe15,Ke13,Ke15}. For Landau quantized EoS, we get stable super-Chandrasekhar white dwarf masses \citep{Mu17}. 

    Table-I shows variations of the masses and radii of non-magnetized white dwarfs with central electron number density $n_e$. Table-II shows the mass-radius relationship of magnetized white dwarfs with no Landau quantization of electrons with $n_e$ for central magnetic field strength $B_d(centre)=0.1$ and fixed surface field $B_s=10^9$ gauss. It is interesting to note that varying the central magnetic field strength from 0.1 to 10 produced no significant changes of masses and radii for non-Landau quantized white dwarfs from the non-magnetized ones at the same central electronic number density. Table-III shows the mass-radius relationship of magnetized white dwarfs with Landau quantization of electrons for the central electronic number density $\sim 5\times10^{-6}$ $fm^{-3}$ (corresponding to the central density of the maximum mass of non-magnetized white dwarf) for $B_d(centre)=0.1,1,2,5,7,10$ and $B_s=10^9$ gauss. Slight variations of the central density is needed due to convergence of the magnetic field strength at the centre. In this case we get super-Chandrasekhar white dwarfs. 
    
\begin{table}[h!]
\centering
\caption{Variations of masses and radii of non-magnetic white dwarfs with central number density of electrons which can be expressed in units of 2$\times10^9$ $g cm^{-3}$ for mass density by multiplying with 1.6717305$\times10^6$.}
\begin{tabular}{||c|c|c||}
\hline 
\hline
$~~~~n_e$ (r=0)$~~~~$&$~~~~$Radius$~~~~$ &$~~~~$Mass$~~~~$ \\ \hline
fm$^{-3}$&Kms & $M_\odot$ \\ \hline
\hline
5.0$\times$10$^{-6}$&1126.44&1.3968 \\
4.0$\times$10$^{-6}$&1202.12&1.3959 \\
3.0$\times$10$^{-6}$&1306.22&1.3942 \\
2.0$\times$10$^{-6}$&1466.17&1.3902 \\
1.0$\times$10$^{-6}$&1778.39&1.3787 \\
 \hline
\hline
\end{tabular}
\label{table1} 
\end{table}
\noindent 

\begin{table}[h!]
\centering
\caption{Variations of masses and radii of magnetized and non-Landau quantized white dwarfs with central number density of electrons which can be expressed in units of 2$\times10^9$ $g cm^{-3}$ for mass density by multiplying with 1.6717305$\times10^6$ for $B_d(centre)=0.1=4.414\times10^{12}$ gauss.}
\begin{tabular}{||c|c|c||}
\hline 
\hline
$~~~~n_e$ (r=0)$~~~~$&$~~~~$Radius$~~~~$ &$~~~~$Mass$~~~~$ \\ \hline
fm$^{-3}$&Kms & $M_\odot$ \\ \hline
\hline
7.0$\times$10$^{-6}$&1020.31&1.3973 \\
6.0$\times$10$^{-6}$&1067.79&1.3972 \\
5.0$\times$10$^{-6}$&1126.47&1.3968 \\
4.0$\times$10$^{-6}$&1202.15&1.3960 \\
3.0$\times$10$^{-6}$&1306.26&1.3942 \\
2.0$\times$10$^{-6}$&1466.22&1.3903 \\
 \hline
\hline
\end{tabular}
\label{table2} 
\end{table}
\noindent 

\begin{table}[htbp]
\centering
\caption{Variations of masses and radii of magnetized and Landau quantized white dwarfs with central number density of electrons ($\sim 5\times10^{-6}$ $fm^{-3}$) which can be expressed in units of 2$\times10^9$ $g cm^{-3}$ for mass density by multiplying with 1.6717305$\times10^6$. The maximum magnetic field $B_{dc}$ at the centre is listed in units of $B_c$ whereas the surface magnetic field $B_s$ is taken to be 10$^{9}$ gauss.}
\scalebox{0.83}{
\begin{tabular}{||c|c|c|c||}
\hline 
\hline
$~~~~n_e$ (r=0)$~~~~$&$~~~~$Radius$~~~~$ &$~~~~$Mass$~~~~$&$~~~~$$B_{dc}$$~~~~$ \\ \hline
fm$^{-3}$&Kms & $M_\odot$ &in units of $B_c$ \\ \hline
\hline
4.674543$\times$10$^{-6}$&1131.48&1.3968&0.1 \\
4.674690$\times$10$^{-6}$&1164.28&1.4074&1.0 \\
4.674209$\times$10$^{-6}$&1349.45&1.4339&2.0 \\
4.670830$\times$10$^{-6}$&1503.64&1.6863&5.0 \\
4.677677$\times$10$^{-6}$&1663.86&2.0217&7.0 \\
4.661657$\times$10$^{-6}$&1954.44&2.8997&10.0 \\ \hline
\hline
\end{tabular}}
\label{table3} 
\end{table}
\noindent 

    For the above mentioned configurations of white dwarfs, we calculate the r-mode instability windows from the fiducial timescales and solving for the critical angular frequency as a function of core temperature and surface magnetic field using Eq.(13). In Figs. 1, 2 and 3 we plot the reduced critical frequency as a function of temperature for non-magnetized, magnetized but not Landau quantized and magnetized and Landau quantized rapidly rotating hot sub- and super-Chandrasekhar white dwarfs of different masses respectively. In Fig. 4 we plot the r-mode instability windows of $1.3968 M_\odot$ white dwarf for all the three cases to show explicitly the effects of Landau quantization and magnetic braking. Table-IV shows the mass, radius, central magnetic field and critical frequency at $10^8$ K of $1.3968 M_\odot$, $2.0217 M_\odot$ and $2.8997 M_\odot$.
    
    In Fig. 5 we plot the time evolution of rotational frequencies of hot white dwarfs by integrating Eq.(15). We take the initial spin frequency to be the Keplerian frequency $\nu_K=\frac{\Omega_K}{2\pi}\approx \frac{1}{2\pi}\sqrt{\frac{GM}{R^3}}=\frac{2}{\sqrt{3}}\nu_0$. We take the amplitude of r-mode $\alpha_r$ for $\Omega=\Omega_0$ and keep it constant for a star of given mass $M$ and radius $R$ as it does not vary much. We take the surface temperature $T_{eff}=10^5$ K. In Fig. 6 we plot the time dependence of spindown rates and in Fig. 7 we plot the dependence of spindown rate on frequency. In these figures we took the same constant mass of $1.3968 M_\odot$ to show the effects of Landau quantization and magnetic braking and also two super-Chandrasekhar white dwarf masses of $2.0217 M_\odot$ and $2.8997 M_\odot$ to show the mass dependence.

From Figs. 1, 2 and 3 we see that hot massive rapidly rotating sub- and super-Chandrasekhar accreting white dwarfs have lower reduced critical frequencies and therefore have a high probability to dwell in r-mode instability region and thus emit gravitational radiation. The reduced critical frequencies are lower for heavier white dwarfs at the same temperature and magnetic field. This is because of the fact that the ratios ${|\widetilde{\tau_{GR}}|}/{\widetilde{\tau_{M}}}$ and ${|\widetilde{\tau_{GR}}|}/{\tau_{SV}}$ rapidly decrease with increase in mass \citep{Mu18}. For the same magnetic field and mass, ${\Omega_c}/{\Omega_0} \propto {T^{-1/3}}$. For the same mass Fig. 4 depicts the effects of magnetic braking and Landau quantization (EoS) on the r-mode instability region. From Fig. 4 we see that magnetic braking increases the r-mode instability regions of magnetized white dwarfs and Landau quantization of electrons added with magnetic braking increases the instability regions even more. At the same temperature, the critical frequency of a Landau quantized white dwarf is lower than that of a non-Landau quantized magnetized white dwarf which is again lower than that of a magnetized white dwarf (see Table-IV). Magnetic fields can maintain the oscillation mode and prevents damping. This makes massive super-Chandrasekhar white dwarfs more prone to gravitational wave emission from r-mode instability which can be detected in the near future.

From Figs. 5, 6 and 7 we conclude that the spindown and spindown rates are higher for magnetized white dwarfs than those of non-magnetized ones because of the fact that the ratio ${\widetilde{\tau_{M}}}/{|\widetilde{\tau_{GR}}|}\propto {B_s^{-2}}$ for a fixed mass. In the saturated phase of the r-mode at later times the magnetic braking timescale becomes less than the gravitational radiation timescale which makes magnetic braking the dominant spindown mechanism. There is no considerable effect of Landau quantization in the spindown and spindown rates.

\begin{table}[htbp]
\centering
\caption{Critical frequencies of magnetized and non-magnetized hot white dwarfs for core temperature of $10^8$ K.}
\scalebox{0.75}{
\begin{tabular}{||c|c|c|c||}
\hline 
\hline
$~~$Mass$~~$&$~~$Radius$~~$&$~~~~$$B_{dc}$$~~~~$&$~~\nu_c(10^8 K)~~$ \\ \hline
$M_\odot$&Kms&in units of $B_c$&Hz \\ \hline
\hline
1.3968&1126.44&0.0&1.1698 \\
1.3968&1126.47&0.1 (no Landau quantization)&1.1582 \\
1.3968&1131.48&0.1 (Landau quantization)&1.1546 \\
2.0217&1663.86&7.0 (Landau quantization)&0.8283 \\
2.8997&1954.44&10.0 (Landau quantization)&0.6335 \\ \hline
\hline
\end{tabular}}
\label{table4} 
\end{table}
\noindent 

\subsection{Gravitational wave strain amplitude}
\label{subsection 3.2} 

    The gravitational wave strain amplitude $h_0^{r-mode}$ from r-modes dominated by $l=m=2$ current quadrupole emission is given by \citep{Ma13} 
    
\begin{equation}
h_0^{r-mode}=\sqrt{\frac{8\pi}{5}}\frac{G}{c^5}\frac{1}{r}\alpha_r\omega_r^3MR^3\tilde{J}
\label{eq53}
\end{equation}
where $r$ is the distance of the white dwarf and $\omega_r=\frac{4}{3}\Omega$ is the angular frequency of the r-mode.

The expressions of $\alpha_r$ (Eq.(50)) and $h_0^{r-mode}$ \\
(Eq.(52)) can be written as \citep{Pa18}

\begin{equation}
\alpha_r=7.9494\times10^{-17}\left[\frac{|\tau_{GR}|}{\tilde{J}}\right]^{1/2}\frac{\sigma^{1/2}T_{eff}^2}{\Omega}\left[\frac{M_\odot}{M}\right]^{1/2}
\label{eq54}
\end{equation}
and
\begin{equation}
h_0^{r-mode}=2.9126\times10^{-16}(\alpha_r\tilde{J})\left[\frac{\Omega}{Hz}\right]^3\left[\frac{R}{km}\right]^2\left[\frac{M}{M_\odot}\right]
\label{eq55}
\end{equation}

\begin{figure}[htbp]
\vspace{0.0cm}
\eject\centerline{\epsfig{file=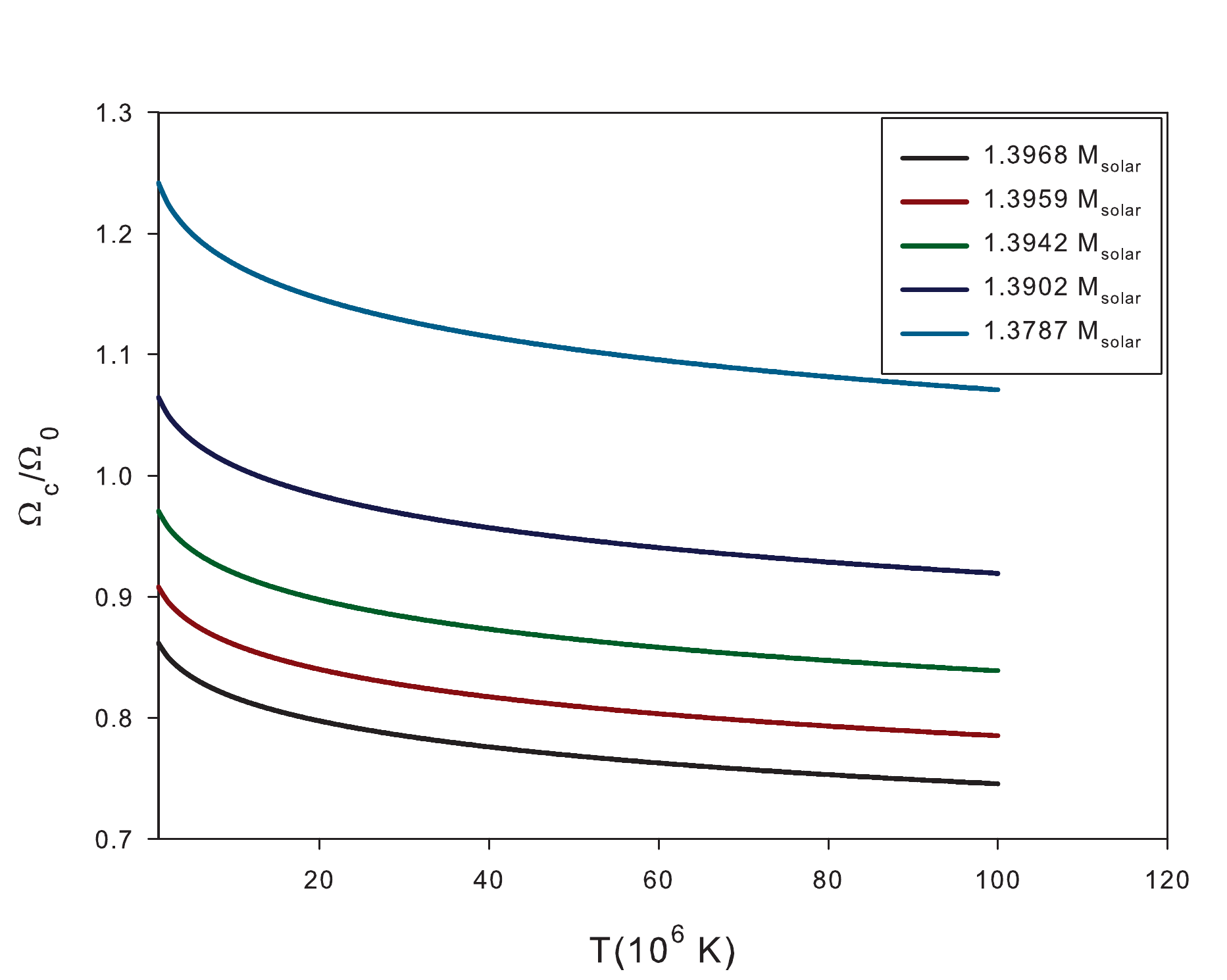,height=8.5cm,width=9cm}}
\caption{Plots of reduced critical angular frequency with temperature for different masses of non-magnetized white dwarfs.} 
\label{fig1}
\vspace{0.0cm}
\end{figure}

\begin{figure}[htbp]
\vspace{0.0cm}
\eject\centerline{\epsfig{file=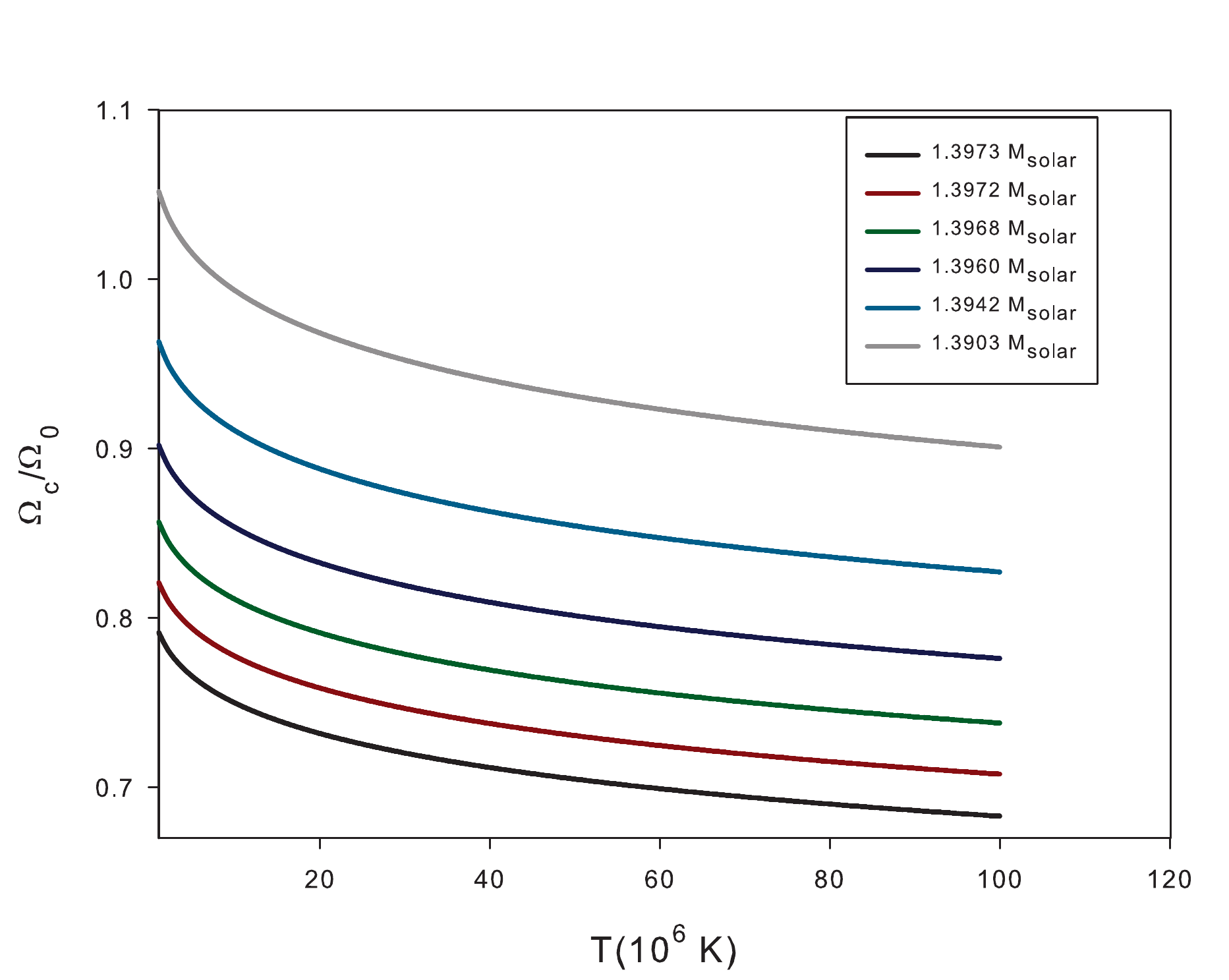,height=8.5cm,width=9cm}}
\caption{Plots of reduced critical angular frequency with temperature for different masses of magnetized white dwarfs with no Landau quantization of electrons. The surface magnetic field is fixed at $10^9$ gauss.} 
\label{fig2}
\vspace{0.0cm}
\end{figure}

\begin{figure}[htbp]
\vspace{0.0cm}
\eject\centerline{\epsfig{file=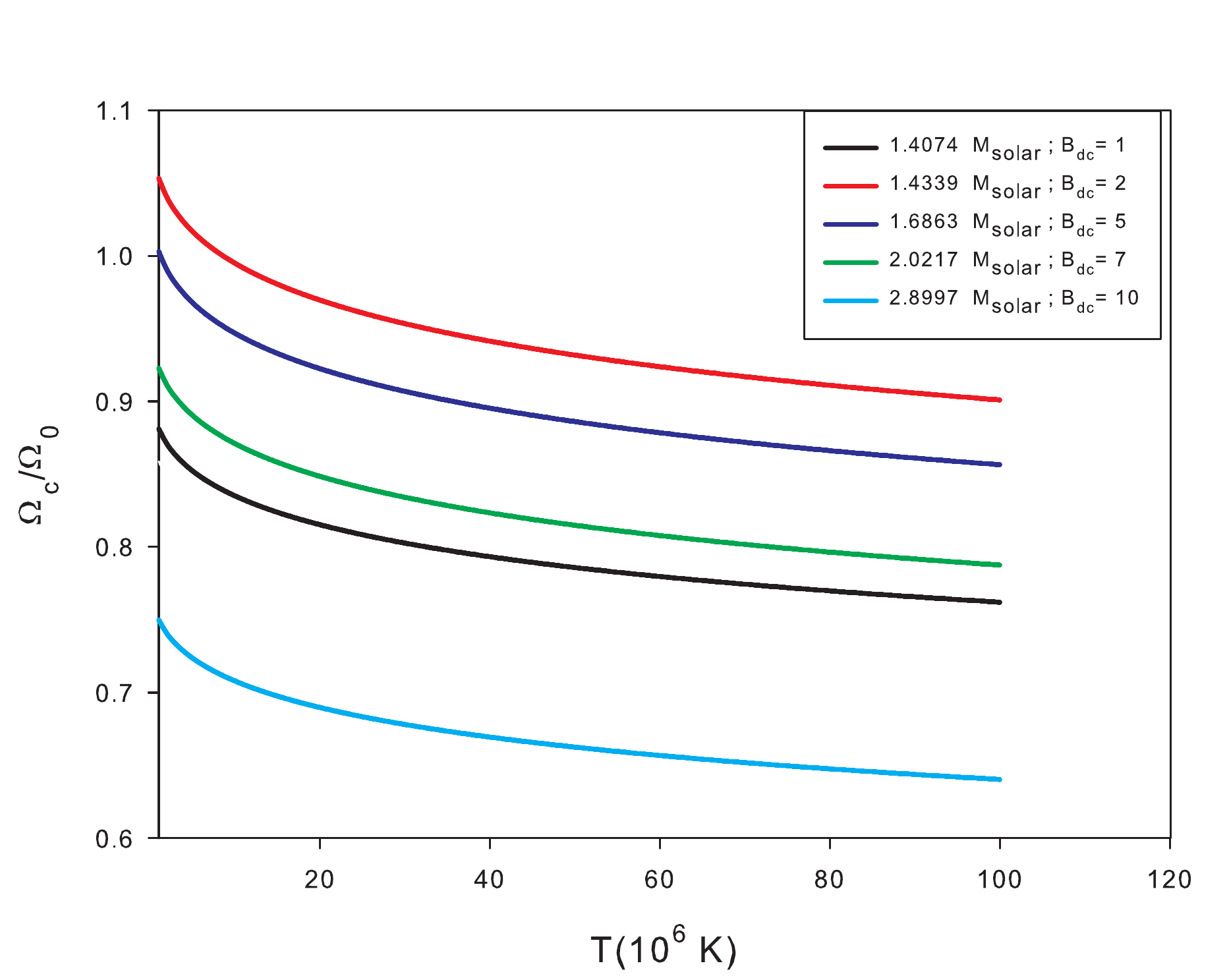,height=8.5cm,width=9cm}}
\caption{Plots of reduced critical angular frequency with temperature for different masses of magnetized white dwarfs with Landau quantization of electrons. The surface magnetic field is fixed at $10^9$ gauss.} 
\label{fig3}
\vspace{0.0cm}
\end{figure}

\begin{figure}[htbp]
\vspace{0.0cm}
\eject\centerline{\epsfig{file=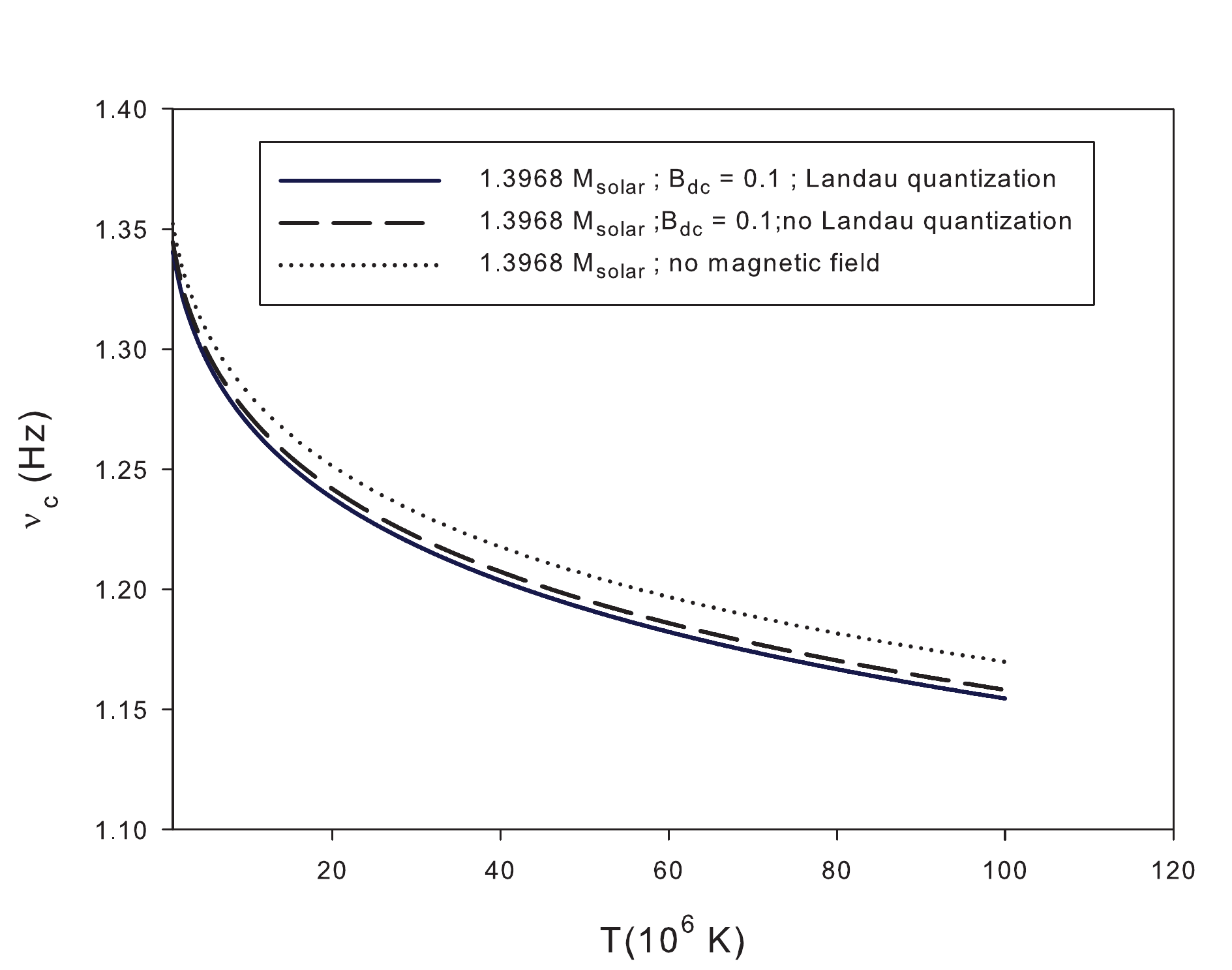,height=8.5cm,width=9cm}}
\caption{Plots of critical angular frequency with temperature for a non-magnetized, magnetized but not Landau quantized and magnetized and Landau quantized white dwarf with mass of $1.3968 M_\odot$ and surface magnetic field of $10^9$ gauss.} 
\label{fig4}
\vspace{0.0cm}
\end{figure}

\begin{figure}[htbp]
\vspace{0.0cm}
\eject\centerline{\epsfig{file=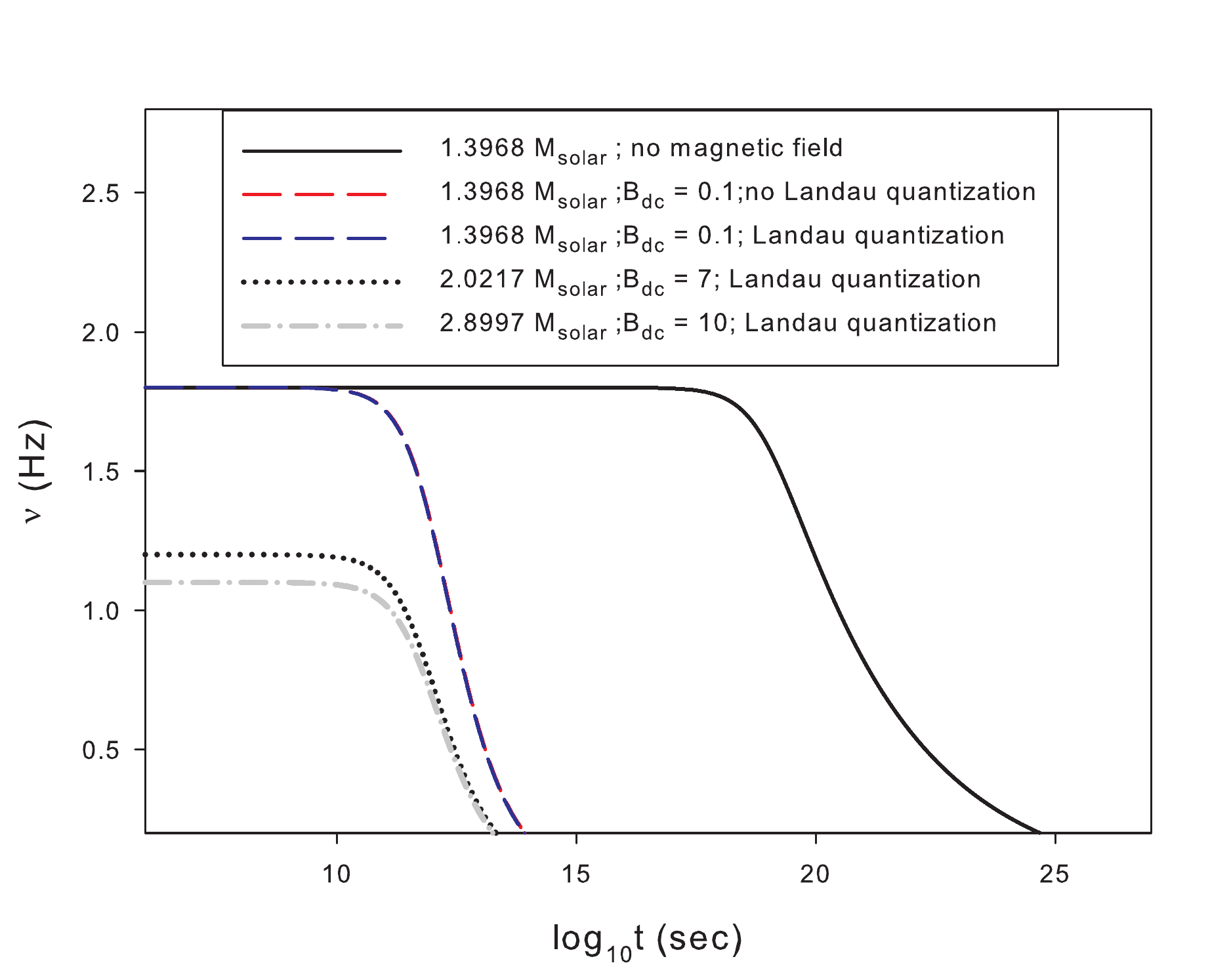,height=8.5cm,width=9cm}}
\caption{Plots of time evolution of rotational frequency of magnetized and non-magnetized white dwarfs of masses of $1.3968 M_\odot$, $2.0217 M_\odot$ and $2.8997 M_\odot$ through gravitational radiation and magnetic braking. } 
\label{fig5}
\vspace{0.0cm}
\end{figure}

\begin{figure}[h!]
\vspace{0.0cm}
\eject\centerline{\epsfig{file=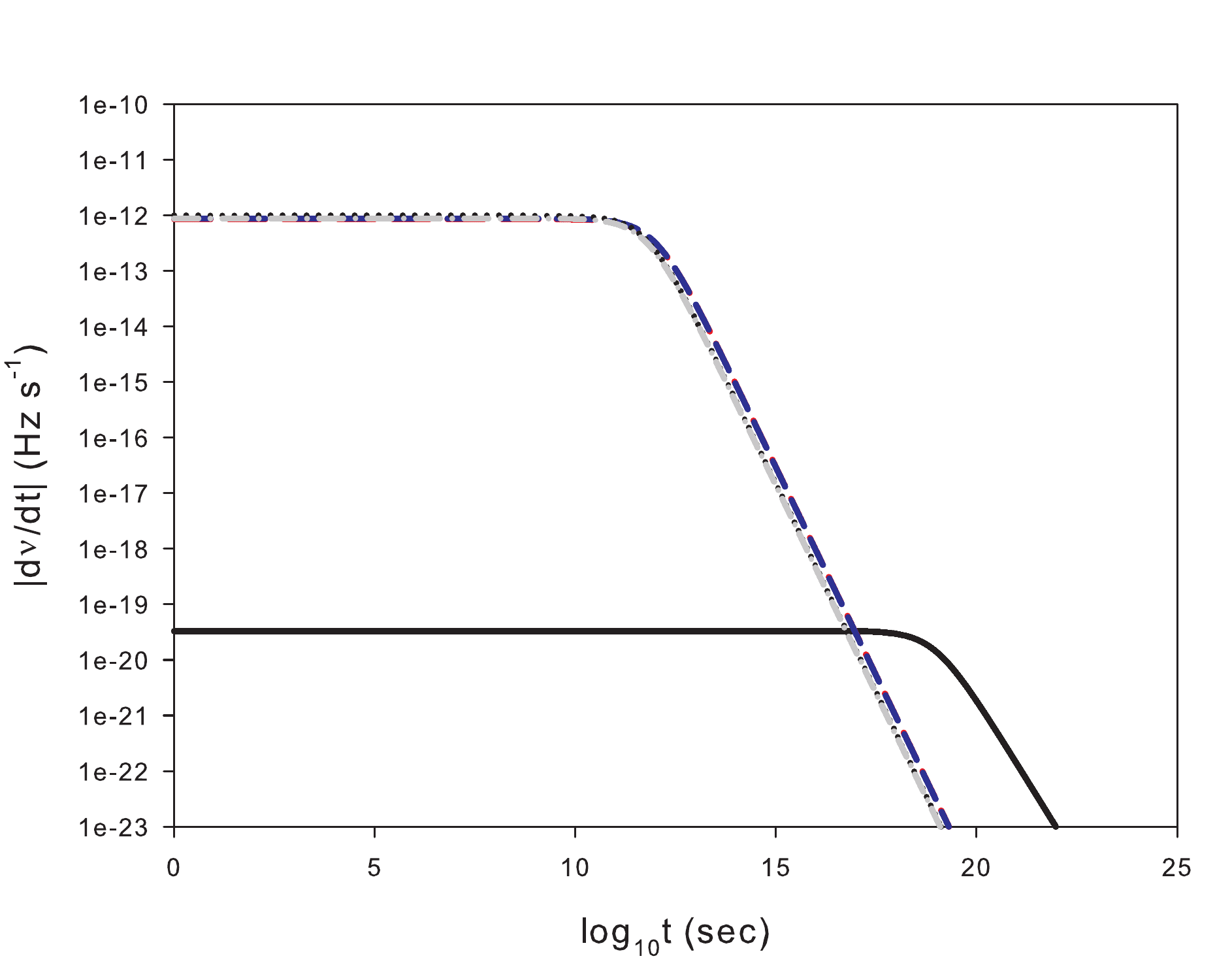,height=8.5cm,width=9cm}}
\caption{Plots of time evolution of spindown rates of magnetized and non-magnetized white dwarfs of masses of $1.3968 M_\odot$, $2.0217 M_\odot$ and $2.8997 M_\odot$ through gravitational radiation and magnetic braking. The legends are the same as in Fig. 5. } 
\label{fig6}
\vspace{0.0cm}
\end{figure}

\begin{figure}[h!]
\vspace{0.0cm}
\eject\centerline{\epsfig{file=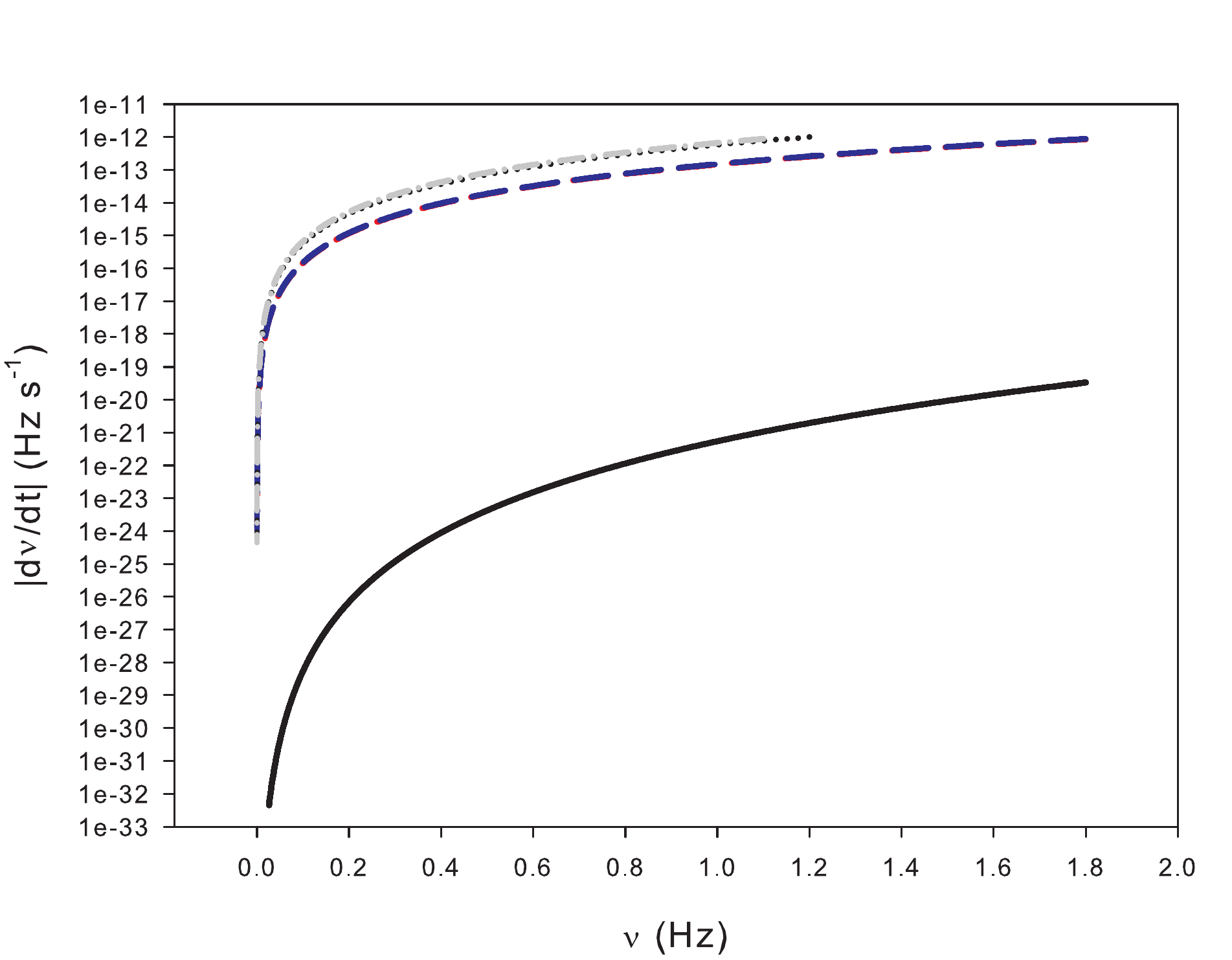,height=8.5cm,width=9cm}}
\caption{Spindown rates as functions of rotational frequencies of magnetized and non-magnetized white dwarfs of masses of $1.3968 M_\odot$, $2.0217 M_\odot$ and $2.8997 M_\odot$ through gravitational radiation and magnetic braking. The legends are the same as in Fig. 5. } 
\label{fig7}
\vspace{0.0cm}
\end{figure}

    Table-V lists the strain amplitudes due to r-mode instability of sub- and super-Chandrasekhar white dwarfs at a distance of 1 kpc rotating at $\nu_0=\frac{\sqrt{3}}{2}\nu_K$. The masses and radii are the same as in Table-IV.

\begin{table}[htbp]
\centering
\caption{Gravitational wave strain amplitudes from r-mode instability of magnetized and non-magnetized sub- and super-Chandrasekhar hot white dwarfs at a distance of 1 kpc.}
\begin{tabular}{||c|c|c||}
\hline 
\hline
$~~$Mass$~~$&$~~$Radius$~~$&$~~$$h_0^{r-mode}$$~~$ \\ \hline
$M_\odot$&Kms&$10^{-27}$ \\ \hline
\hline
1.3968&1126.44&3.6156 \\
1.3968&1126.47&3.6158 \\
1.3968&1131.48&3.6561 \\
2.0217&1663.86&6.6084 \\
2.8997&1954.44&6.8704 \\ \hline
\hline
\end{tabular}
\label{table5} 
\end{table}
\noindent 

\section{Summary and Conclusions}
\label{Section 4}

    In the present study we have investigated the r-mode instability regions of young and accreting sub- and super-Chandrasekhar white dwarfs in presence of magnetic braking and Landau quantization of the electron gas in the EoS with the surface magnetic field fixed at $\sim 10^9$ gauss. We have taken the white dwarfs to be thermally equilibrated and thereby constrained the saturated r-mode amplitudes. We calculated the critical frequencies as functions of the core temperatures of three different categories of white dwarfs: non-magnetized, magnetized but not Landau quantized and magnetized and Landau quantized sub- and super-Chandrasekhar white dwarfs. We found that at the same core temperature, the critical frequency of the Landau quantized white dwarf is the least, followed by that of the non-Landau quantized one. The critical frequency of the non-magnetized one is the highest. Hence, the r-mode instability window is the greatest for Landau quantized and magnetized white dwarfs at high temperatures. Landau quantization increases the instability window. Both Landau quantization and magnetic braking try to keep the mode unstable and prevent damping. We have also found the spindown and spindown rates under the combined effects of gravitational radiation and magnetic braking. We saw that in the saturated phase at later stages of evolution, magnetic braking dominates gravitational radiation and hence magnetic white dwarfs spin down rapidly. The spindown rates of magnetic white dwarfs are also high. Finally, we computed the gravitational wave strain amplitudes from r-mode instability of the white dwarfs at a distance of 1 kpc. We conclude that amplitudes due to r-mode instability are $\sim 10^{-27}$, which make white dwarfs excellent candidates for detection of gravitational waves in future. These detections may provide clues to the nature of high density condensed matter and phase transitions in white dwarf interiors, existence of super-Chandrasekhar masses and white dwarf pulsars and also on the nature of gravity.

\newpage    
\section{Acknowledgements} 
\label{Section 5} 

    The authors would like to thank DAE-BRNS, India for the research grant.

%

\end{document}